\documentclass{elsart}

\usepackage{epsfig}
\usepackage{amsmath}
\input amssym

\begin{document}

\begin{frontmatter}



\title{Energy conservation and scaling violations in particle production}

\author{J.~Dias de Deus and J.~G.~Milhano}

\address{CENTRA, Instituto Superior T\'ecnico (IST), Av. Rovisco Pais, P-1049-001 Lisboa, Portugal}

\begin{abstract}
We use a simple Colour Glass Condensate/String Percolation Model argument to show the existence, due to energy conservation, of bounds to the violation of Feynman scaling and limiting fragmentation.
\end{abstract}


\end{frontmatter}

Experimental studies of inclusive particle distributions in the 1960s led to the introduction of two scaling laws --- Feynman scaling \cite{Feynman:1969ej} and limiting fragmentation \cite{Benecke:1969sh} --- concerning the longitudinal momentum fraction $x$ or rapidity $y$. 
Feynman scaling refers to the small-$x$ ($x\simeq 0$) region, while limiting fragmentation concerns the region $y\simeq Y$, where $Y$ is the beam rapidity.
In the present paper, we do not aim to discuss the validity of the above mentioned scaling laws, but rather to address the consequences of their asymptotic, i.e. as $Y\rightarrow \infty$, violation.

Feynman scaling concerns the production of slow partons (gluons) and states, in analogy with photon radiation in QED, the phase space independence of the produced density.
If we define the transverse momentum $p_T$ integrated density $\rho (y, Y)$ as
\begin{equation} 
\label{eq:rhodef}
	\rho (y, Y) \equiv \frac{dn}{dy}\, ,
\end{equation}
where $y$ is the rapidity of the produced particle  and $Y$ is the beam rapidity, Feynman scaling means that\footnote{Note that the limit  $y=const\, ;\,\,\,Y\rightarrow \infty$ corresponds to $x\rightarrow 0$.}
\begin{equation}
\label{eq:feynscal} 
	\rho(y, Y) \xrightarrow[y=const\, ,\,\,\,Y\rightarrow \infty]{} const\, .
\end{equation}

On the other hand, in relation to limiting fragmentation, the interest is focused on large rapidities $y\sim Y$. Limiting fragmentation essentially states that the $p_T$-integrated distribution of fragments  of the beam particle seen from the target is independent of the energy of the collision. If we treat $\rho$ as a function  of $\Delta \equiv y-Y$ and $Y$, rather than as a function of $y$ and $Y$,  limiting fragmentation amounts to 
\begin{equation} 
\label{eq:lf} 
	\rho(\Delta, Y) \xrightarrow[\Delta=const\, ,\,\,\,Y\rightarrow \infty]{} f(\Delta)\, .
\end{equation}

Conservation of energy can be written in the form
\begin{equation}
\label{eq:encons}
	\sum_{\{ i \}} \int_0^{Y_i} \langle m_T^i\rangle \cosh y_i \,\rho (y_i,Y)\, dy_i = \frac{N_{part}}{2} \frac{\sqrt{s}}{2}\, ,
\end{equation}
where $N_{part}$ is the number of participating nucleons (we are assuming symmetric AA collisions), $\{i\}$ is  the set of particles in  the final state $(i=\pi^{+}$, $\pi^{-}$, $\bar{p}$, $K^+$, $\ldots )$, $y_i$ is the rapidity of the $i$-th final state particle, and $ \langle m_T^i\rangle$ is its transverse mass $ \langle \sqrt{{m^i}^2 +{p_T^i}^2}\rangle$.

In order to simplify the problem we shall fix the species in the final state ($\pi^+$ or $\bar{p}$, for instance) and limit the region of integration, in general ranging from zero to $Y$, to some convenient region $\{ R\}$, whether that is the Feynman scaling region  $\{ FS\}$ or the limiting fragmentation region $\{ LF\}$. As a result, we have, from (\ref{eq:encons}), 
\begin{equation}
\label{eq:restencons}
	\int_{\{ R\}} \frac{\langle m_T\rangle}{m} \cosh y\, \rho (y,Y)\, dy < \frac{N_{part}}{2} \frac{e^Y}{2}\, ,
\end{equation}
where $m$ is the mass of the beam particle.

In order to test the fulfilment of  (\ref{eq:restencons}) we require information on $\rho(y,Y)$ ---  which we have in both the cases of Feynman scaling and limiting fragmentation --- and information on $\langle m_T(y,Y)\rangle$ (or $\langle m_T(\Delta,Y)\rangle$). As in  (\ref{eq:restencons}) we have fixed one species in the final state, the problem is reduced to studying the dependence of $\langle p_T^2(y,Y)\rangle$ (or $\langle p_T^2(\Delta,Y)\rangle$) on $Y$.

In approaches that account for saturation phenomena, namely the Colour Glass Condensate \cite{McLerran:1993ni,McLerran:1993ka,McLerran:1994vd,Jalilian-Marian:1996xn,Jalilian-Marian:1997dw,McLerran:2001cv,SchaffnerBielich:2001qj} (where the saturation momentum sets the scale for other physical quantities) and the String Percolation Model \cite{Armesto:1996kt,Nardi:1998qb,Dias de Deus:2003fg,Dias de Deus:2005yt} (where Gauss' theorem links $\rho$ to $\langle p_T^2 \rangle$) there is a simple relation between $dn/dy$ and $\langle p_T^2 \rangle$, 
\begin{equation} 
\label{eq:ptrho}
	\langle p_T^2 \rangle \sim \frac{1}{S_A} \rho\, ,
\end{equation}
where $S_A\sim A^{2/3}$ is the size of the overlapping area in the collision.
Essentially, what (\ref{eq:ptrho}) tells us is that $\rho$ and $\langle p_T^2 \rangle$ as functions of the total rapidity $Y$ have identical behaviour.
 If $\rho$ increases with $Y$ or centrality, then $\langle p_T^2 \rangle$ will also increase with $Y$ or centrality.

We now turn to consider Feynman scaling. If scaling (\ref{eq:feynscal}) is satisfied then  $\langle m_T \rangle$, given (\ref{eq:ptrho}), is constant and (\ref{eq:restencons}) is trivially satisfied.\footnote{Note that $\{ R\}\equiv\{ FS\}$ is a finite region in $y$.}

It is, however, well known that Feynman scaling is violated \cite{Alper:1975jm}. Gluons, unlike photons, are self-interacting and their number increases exponentially with phase-space (see, for instance \cite{Kharzeev:2001gp}). Thus, we expect
\begin{equation}
\label{eq:feynscalviol} 
	\rho(y, Y) \xrightarrow[y=const\, ,\,\,\,Y\rightarrow \infty]{} e^{\lambda Y}\, ,
\end{equation}
with $0\leq \lambda \leq 1$.
Using (\ref{eq:feynscalviol}) in (\ref{eq:restencons}), with (\ref{eq:ptrho}), we obtain the bound 
\begin{equation}
\label{eq:lambdabound} 
	\lambda\leq \frac{2}{3}\, .
\end{equation} 
The value of $\lambda \simeq 0.288$, phenomenologically extracted \cite{GolecBiernat:1998js} from experimental data is obviously within this bound.

The experimental situation regarding limiting fragmentation is not very clear.
The rather limited range of $Y$, from 3 to 5.3, for which experimental data is available allied with the intrinsic smallness of $\rho$ in the limiting fragmentation $y\lesssim Y$ region perclude, at present, definite statements regarding the occurence of limiting fragmentation.
For pions, limiting fragmentation seems to hold \cite{Adams:2005cy}, while it is certainly not satisfied for proton production  \cite{Adams:2005cy}.
For charged particle production the situation is ambiguous, but detailed studies of RHIC data \cite{Adams:2005cy,Brogueira:2006nz} indicate that $\rho$, at fixed $\Delta$, is decreasing with increasing $Y$. This is, in a sense, what would be expected from the decrease of $dn/dy$ for protons at large $y$.

In order to address  limiting fragmentation it is convenient to change from the variable $y$ to $\Delta$ and rewrite (\ref{eq:restencons}) in the form
\begin{equation}
\label{eq:restenconsdelta}
	\int_{\{ R\}} \frac{\langle m_T\rangle}{m}  \rho (\Delta,Y) \, e^\Delta\,  d\Delta < \frac{N_{part}}{2} \, ,
\end{equation}
where now $\{ R\}\equiv \{ LF\}$ refers to a  $Y$-independent finite region in  $\Delta$. If limiting fragmentation occurs (\ref{eq:lf}), $\rho$ is independent of $Y$ and, from (\ref{eq:ptrho}), $\langle p_T^2 \rangle$ is also independent of $Y$ and (\ref{eq:restenconsdelta}) can, in principle, be satisfied. It should noted that the limit $Y\rightarrow\infty$ does not affect this result.

At this point it is natural to consider what happens when limiting fragmentation is \textit{not} satisfied.
If we take, as an example, the simple model of \cite{DiasdeDeus:2007wb}, where the rapidity extended  colour fields (alternatively, valence strings) describing the system immediately after the collision are the sources for particle production, one arrives at the plausible evolution equation 
\begin{equation}
\label{eq:log}
	\frac{\partial \rho}{\partial (-\Delta)} = \frac{1}{\delta}\, \rho\,\Big(1 - \frac{\rho}{\rho_Y}\Big)\, ,
\end{equation}
where $\delta$ is a constant that controls the low density $\rho$ evolution and $\rho_Y$ is the, increasing with total rapidity $Y$, limiting value of $\rho$ saturation. In the linear approximation limit, i.e. when $\rho$ is small, we obtain the solution 
\begin{equation}
\label{eq:rhosol}
	 \rho(\Delta,Y) ={\cal C}(Y) e^{-\Delta/\delta}\, .
\end{equation}
If ${\cal C}(Y) = const.$ we recover the limiting fragmentation limit.

\begin{figure}[h] 
   \centering
   \includegraphics[angle=0,width=10cm]{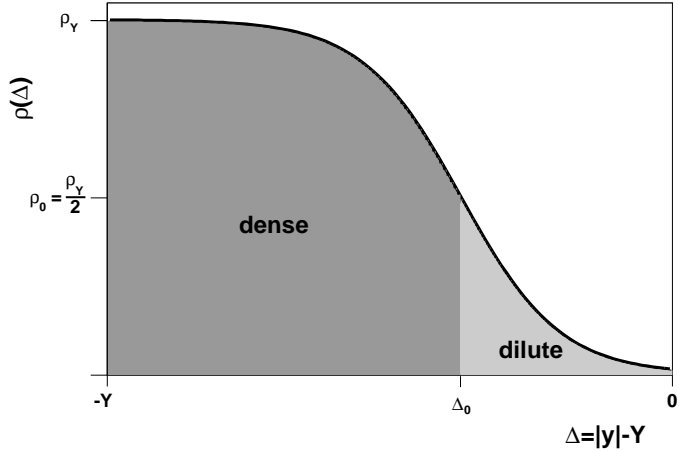} 
   \caption{The distribution (\ref{eq:rhoint}) showing saturation at large $-\Delta$, the change of curvature at the point $\Delta_0 = -\alpha Y$ which separates between the dense ($\Delta < \Delta_0$) and the dilute ($\Delta > \Delta_0$) regions.}
   \label{fig:solution}
\end{figure}

In general (for an asymmetrical situation see \cite{DiasdeDeus:2007wb}) the solution of (\ref{eq:log}) is given by
\begin{equation}\label{eq:rhoint}
	\rho( \Delta, Y) = \frac{e^{\lambda Y}}{e^{\frac{\Delta +\alpha Y}{\delta}}+1}\, ,
\end{equation}
where $\lambda$, $\alpha$ and $\delta$ are constants: $\lambda$, as seen before, controls the rise of the $Y\simeq 0$ plateau, $\alpha$ the transition from the dilute to the dense region and $\delta$ the slope of that transition. Eq. (\ref{eq:log}) is illustrated in 
Fig.\,\ref{fig:solution}.

A fit, performed in \cite{Brogueira:2006nz}, of  (\ref{eq:log}) to RHIC data yields the following parameter values: $\lambda = 0.247$ which is both within the Feynman scaling bound (\ref{eq:lambdabound}) and consistent with the phenomenological extraction of 
\cite{GolecBiernat:1998js}, $\alpha = 0.269$,  and $\delta = 0.67$. The model of  \cite{DiasdeDeus:2007wb}, with the parameters fixed by RHIC data, was used to predict 
multiplicity distributions in rapidity for Pb-Pb and p-Pb at LHC energies in \cite{Abreu:2007zw,Abreu:2007kv}.

In the limiting fragmentation limit (\ref{eq:lf}) we obtain, from (\ref{eq:rhoint}),
\begin{equation} 
\label{eq:lfsol} 
	\rho(\Delta, Y) \xrightarrow[\Delta=const\, ,\,\,\,Y\rightarrow \infty]{} e^{(\lambda -\frac{\alpha}{\delta}) Y} e^{-\frac{\Delta}{\delta}}\, .
\end{equation}

Note that  (\ref{eq:lf}) is simply a realization of (\ref{eq:rhosol}). The condition for limiting fragmentation is thus given by
\begin{equation} 
\label{eq:lfcond} 
	\Big(\lambda -\frac{\alpha}{\delta}\Big) = 0\, ,
\end{equation}
and violation of limiting fragmentation amounts to non-fulfillment of this equality.
The behaviour of $\rho$, for fixed $\Delta$, as a function of $Y$ depends on the sign of $(\lambda -\frac{\alpha}{\delta})$. 
If $(\lambda -\frac{\alpha}{\delta}) > 0$, $\rho$ increases indefinitely with increasing $Y$, the same happens for $\langle p_T^2 \rangle$, and the energy conservation bound (\ref{eq:restenconsdelta}) is violated.
If, however,  $(\lambda -\frac{\alpha}{\delta}) < 0$,  $\rho$ decreases  indefinitely with increasing $Y$,  $\langle p_T^2 \rangle$ also decreases, and the energy conservation bound  (\ref{eq:restenconsdelta}) is satisfied, asymptotically, in a trivial way with the left hand side of  (\ref{eq:restenconsdelta}) being zero.

If this is indeed what is happening, i.e the contribution to energy conservation of fast particles, presumably nucleons carrying energy of the order $e^Y$, is asymptotically vanishing, then the relevant question to be asked is: from where do the contributions needed to saturate energy conservation come?

String percolation theory offers a simple answer. Percolation implies, on one hand, a limited rise of the dense region plateau with $\rho \sim N_{part}$, and, on the other, the generation of higher longitudinal momentum strings and very fast produced particles. The fastest particles will no longer be the fast particles from valence strings (net baryons)  but rather the mesons and baryons arising from sea string percolation \cite{Dias de Deus:2005sq}.

In conclusion, we have presented very simple energy conservation tests to violation of Feynman scaling and of limiting fragmentation. The main conclusions are that an exponential violation of Feynman scaling, $\e^{\lambda Y}$, is only possible for $\lambda\le 2/3$ (\ref{eq:lambdabound}), and that an indefinitely growth of $\rho (\Delta, Y)$, at fixed $\Delta$, in violation of limiting fragmentation, is excluded by the energy conservation bound (\ref{eq:restenconsdelta}). Limiting fragmentation can, however, be violated provided that $\rho$, at fixed $\Delta$, decreses with increasing $Y$.


\begin{thebibliography}{99}

\bibitem{Feynman:1969ej}
  R.~P.~Feynman,
  Phys.\ Rev.\ Lett.\  {\bf 23} (1969) 1415.

\bibitem{Benecke:1969sh}
  J.~Benecke, T.~T.~Chou, C.~N.~Yang and E.~Yen,
  Phys.\ Rev.\  {\bf 188} (1969) 2159.
  
  
  \bibitem{McLerran:1993ni}
  L.~D.~McLerran and R.~Venugopalan,
  Phys.\ Rev.\ D {\bf 49} (1994) 2233
  [arXiv:hep-ph/9309289].

\bibitem{McLerran:1993ka}
  L.~D.~McLerran and R.~Venugopalan,
  Phys.\ Rev.\ D {\bf 49} (1994) 3352
  [arXiv:hep-ph/9311205].
  

\bibitem{McLerran:1994vd}
  L.~D.~McLerran and R.~Venugopalan,
  Phys.\ Rev.\ D {\bf 50} (1994) 2225
  [arXiv:hep-ph/9402335].
  
  \bibitem{Jalilian-Marian:1996xn}
  J.~Jalilian-Marian, A.~Kovner, L.~D.~McLerran and H.~Weigert,
  Phys.\ Rev.\ D {\bf 55} (1997) 5414
  [arXiv:hep-ph/9606337].
  
  
   \bibitem{Jalilian-Marian:1997dw}
  J.~Jalilian-Marian, A.~Kovner and H.~Weigert,
  Phys.\ Rev.\ D {\bf 59} (1999) 014015
  [arXiv:hep-ph/9709432].
  
\bibitem{McLerran:2001cv}
  L.~D.~McLerran and J.~Schaffner-Bielich,
  Phys.\ Lett.\  B {\bf 514} (2001) 29
  [arXiv:hep-ph/0101133].
  
  
\bibitem{SchaffnerBielich:2001qj}
  J.~Schaffner-Bielich, D.~Kharzeev, L.~D.~McLerran and R.~Venugopalan,
  Nucl.\ Phys.\  A {\bf 705} (2002) 494
  [arXiv:nucl-th/0108048].
  
  
\bibitem{Armesto:1996kt}
  N.~Armesto, M.~A.~Braun, E.~G.~Ferreiro and C.~Pajares,
  Phys.\ Rev.\ Lett.\  {\bf 77} (1996) 3736
  [arXiv:hep-ph/9607239].
  
\bibitem{Nardi:1998qb}
  M.~Nardi and H.~Satz,
  Phys.\ Lett.\  B {\bf 442} (1998) 14
  [arXiv:hep-ph/9805247].
 
\bibitem{Dias de Deus:2003fg}
  J.~Dias de Deus, E.~G.~Ferreiro, C.~Pajares and R.~Ugoccioni,
  Phys.\ Lett.\  B {\bf 581} (2004) 156
  [arXiv:hep-ph/0303220].
 
\bibitem{Dias de Deus:2005yt}
  J.~Dias de Deus and R.~Ugoccioni,
  Eur.\ Phys.\ J.\  C {\bf 43} (2005) 249.
 
\bibitem{Alper:1975jm}
  B.~Alper {\it et al.}  [British-Scandinavian Collaboration],
  Nucl.\ Phys.\  B {\bf 100} (1975) 237.
  
\bibitem{Kharzeev:2001gp}
  D.~Kharzeev and E.~Levin,
  Phys.\ Lett.\  B {\bf 523} (2001) 79
  [arXiv:nucl-th/0108006].
  
  
\bibitem{GolecBiernat:1998js}
  K.~J.~Golec-Biernat and M.~Wusthoff,
  Phys.\ Rev.\  D {\bf 59} (1999) 014017
  [arXiv:hep-ph/9807513].
 
 
\bibitem{DiasdeDeus:2007wb}
  J.~Dias de Deus and J.~G.~Milhano,
  Nucl.\ Phys.\  A {\bf 795} (2007) 98
  [arXiv:hep-ph/0701215].
  
\bibitem{Abreu:2007zw}
  S.~Abreu, J.~Dias de Deus and J.~G.~Milhano,
  arXiv:0707.0178 [hep-ph].
  
\bibitem{Abreu:2007kv}
  S.~Abreu {\it et al.},
  arXiv:0711.0974 [hep-ph].
  
  
\bibitem{Adams:2005cy}
  J.~Adams {\it et al.}  [STAR Collaboration],
  Phys.\ Rev.\  C {\bf 73} (2006) 034906
  [arXiv:nucl-ex/0511026].
 
 
\bibitem{Brogueira:2006nz}
  P.~Brogueira, J.~Dias de Deus and C.~Pajares,
  Phys.\ Rev.\  C {\bf 75} (2007) 054908
  [arXiv:hep-ph/0605148].
  
  
\bibitem{Dias de Deus:2005sq}
  J.~Dias de Deus, M.~C.~Espirito Santo, M.~Pimenta and C.~Pajares,
  Phys.\ Rev.\ Lett.\  {\bf 96}, 162001 (2006)
  [arXiv:hep-ph/0507227].
 
 \end{thebibliography}
\end{document}